\pdfoutput=1
\documentclass[11pt]{article}

\usepackage[preprint]{acl}
\usepackage{xspace}
\usepackage{amsfonts}
\usepackage{times}
\usepackage[most]{tcolorbox} 
\usepackage{latexsym}
\usepackage{amsmath} 
\usepackage[T1]{fontenc}
\usepackage[normalem]{ulem}

\usepackage[utf8]{inputenc}
\usepackage{booktabs}
\usepackage{multirow}
\usepackage{colortbl}
\usepackage{tabularx}
\definecolor{mygray}{gray}{0.9}
\usepackage{makecell}
\usepackage{pifont} 

\usepackage{tikz}

\usepackage{microtype}

\usepackage{inconsolata}

\usepackage{graphicx}

\newcommand{\ours}{{\textsc{GPR}\xspace}}
\title{GPR: Empowering Generation with Graph-Pretrained Retriever}

\author{Xiaochen Wang, Zongyu Wu, Yuan Zhong, Xiang Zhang, Suhang Wang, Fenglong Ma\\
The Pennsylvania State University\\
\texttt{\{xcwang,zongyuwu,yfz5556,xzz89,szw494,fenglong\}@psu.edu}
}

\begin{document}
\maketitle
\begin{abstract}
Graph retrieval-augmented generation (GRAG) places high demands on graph-specific retrievers. However, existing retrievers often rely on language models pretrained on plain text, limiting their effectiveness due to \textit{domain misalignment} and \textit{structure ignorance}. To address these challenges, we propose {\ours}, a graph-based retriever pretrained directly on knowledge graphs. {\ours} aligns natural language questions with relevant subgraphs through LLM-guided graph augmentation and employs a structure-aware objective to learn fine-grained retrieval strategies. Experiments on \textbf{two} datasets, \textbf{three} LLM backbones, and \textbf{five} baselines show that {\ours} consistently improves both retrieval quality and downstream generation, demonstrating its effectiveness as a robust retrieval solution for GRAG.

\end{abstract}

\section{Introduction}
\label{sec:intro}

Graph Retrieval-Augmented Generation (GRAG) has emerged as an effective paradigm for enhancing the capabilities of large language models (LLMs)~\cite{min2019knowledge}. By retrieving structured and high-quality knowledge from graphs, these models are able to acquire comprehensive context regarding questions and generate more accurate and grounded responses~\cite{zhang2025survey}.

The effectiveness of GRAG hinges on the quality of the retrieved graph components, placing high demands on the retriever. To meet this challenge, retrievers based on pretrained language models (PLMs)~\cite{karpukhin2020dense} have emerged as a promising solution. These retrievers operate directly on natural language queries without relying on handcrafted rules~\cite{mavromatis2024gnn} or task-specific features~\cite{luo2023reasoning}, offering greater flexibility and generalizability compared to traditional approaches such as non-parametric or graph neural network-based retrievers~\cite{peng2024graph, li2023graph}. However, despite these advantages, existing PLM-based retrievers exhibit the following shortcomings:


\textbf{S1: Domain Misalignment}. Most of the current PLM-based retrievers are built on models pretrained solely on plain text~\cite{he2024g, li2023graph}. These models are proficient in understanding natural language queries, but struggle to interpret graph-structured data, which are composed of semi-structured triplets with irregular formats. The misalignment between query representations in text and the structured nature of graph data leads to suboptimal retrieval, constraining the overall effectiveness of GRAG systems. 

\textbf{S2: Structure Ignorance}. In addition, many approaches directly apply language models to retrieve individual nodes~\cite{he2024g}, triplets~\cite{li2023graph}, or subgraphs~\cite{li2024simple, hu2024grag}, mirroring strategies used in traditional text-based retrieval-augmented generation~\cite{karpukhin2020dense}. However, this overlooks the fundamental property of knowledge graphs: connectivity. Encoding graph elements as isolated units fails to capture the relational structure essential for effective graph retrieval.

To address these limitations, we propose \textbf{G}raph \textbf{P}retrained \textbf{R}etriever ({\ours}), a simple yet effective retriever pretrained directly on knowledge graphs. To resolve \textbf{S1}, {\ours} leverages LLM-guided graph augmentation to align natural language questions with relevant subgraphs, without relying on additional supervision or schema-specific features. To tackle \textbf{S2}, {\ours} employs a structure-aware pretraining objective that distinguishes triplets based on their relevance with questions, encouraging the model to selectively capture comprehensive context that could boost the LLMs.

We evaluate {\ours} on \textbf{two} benchmark datasets using \textbf{three} backbone LLMs and \textbf{five} baselines. Results show that {\ours} consistently retrieves more relevant knowledge from graph and enhances downstream generation. These findings establish {\ours} as a generalizable and effective solution for graph-based retrieval, advancing the development of knowledge-grounded language models.

\section{Related Work}

\label{sec:related_work}
Retrieval-augmented generation (RAG)~\citep{gao2023retrieval,guo-etal-2023-prompt,ma-etal-2023-query} has emerged as a promising approach to mitigate intrinsic limitations of large language models, such as hallucinations~\citep{zhang2023siren,tonmoy2024comprehensive}. Its specialized variant, Graph RAG (GRAG)~\cite{min2019knowledge}, extends this paradigm by retrieving high-quality knowledge from structured knowledge graphs, demonstrating strong potential in knowledge-intensive tasks~\citep{zhang2025survey}.
Pretrained Language Model (PLM)-based retrievers~\cite{karpukhin2020dense} have been widely adopted in GRAG systems, enabling knowledge retrieval at various granularities, including nodes~\cite{he2024g}, triplets~\cite{li2023graph}, and subgraphs~\cite{li2024simple, hu2024grag}. In these approaches, knowledge is encoded using language models pretrained on general plain text, and the retrieved results are either directly fed into the large language models for reasoning or processed by additional adaptation modules to enhance the model’s ability to interpret the retrieved content. While most methods simply leverage pretrained language models, some prior work~\cite{dong2023bridging} has explored pretraining these models on knowledge graphs using conventional objectives such as InfoNCE~\cite{oord2018representation} and Masked Language Modeling~\cite{devlin2019bert}. Nevertheless, these efforts overlook the alignment between the textual query modality and the graph-structured knowledge, often degrading retrieval effectiveness.


\section{\textbf{G}raph \textbf{P}retrained \textbf{R}etriever (GPR)}
\label{sec:method}

\begin{table*}[t]
\centering
\caption{Question answering results (\%) on WebQSP and CWQ datasets. The best-performing results are highlighted in \textbf{bold}.}
\vspace{-10pt}
\label{tab:main_experiments}
\resizebox{\textwidth}{!}{
\begin{tabular}{l|cccc|cccc}
\toprule
\multirow{2}{*}{\textbf{Methods}} & \multicolumn{4}{c|}{\textbf{WebQSP}} & \multicolumn{4}{c}{\textbf{CWQ}} \\ \cline{2-9}
 & \textbf{Accuracy} & \textbf{Precision} & \textbf{Recall} & \textbf{F1} & \textbf{Accuracy} & \textbf{Precision} & \textbf{Recall} & \textbf{F1} \\ 
\midrule
ChatGPT &46.62  &65.97  & 39.48 &49.40  &37.88  &42.19  & 35.63 &38.63  \\
{\:} + G-Retriever & 42.70 & 63.32 &36.44 &46.26 & 33.51 & 39.42 & 31.84 &34.99 \\
{\:} + G-RAG &39.40  &57.55  &33.63 &42.45 &31.87  &36.44  &30.55  &33.21  \\
{\:} + Hybrid & 50.38 & 64.25 & 37.52 &47.37& 39.55 & 44.52 & 37.06 &40.47 \\
{\:} + SKP & 44.80 & 62.10 &36.98 &46.36 &33.43  & 38.71 & 31.71 &34.83\\
\midrule
{\:} + Two Tower & 39.14 & 57.30 & 34.30 &42.67 & 30.85 & 36.34 & 29.60 &32.64 \\
 \rowcolor{mygray} 
{\:} + {\ours} & \textbf{62.40} & \textbf{73.46} & \textbf{46.31} & \textbf{56.79} & \textbf{43.25} & \textbf{47.59} & \textbf{39.59} & \textbf{43.22} \\
\midrule
\midrule
LLaMA2-Chat-7B & 40.16 & 59.82 & 34.85 &43.76 & 28.23 & 9.99 & 28.23 &14.70 \\
{\:}+ G-Retriever & 44.00 & 66.22 &37.68 &47.84 &30.77  &36.51  & 29.32 &32.52  \\
{\:} + G-RAG &20.81  &36.43  &19.29 &25.26 & 9.53 & 11.81 & 9.21 &10.31 \\
{\:} + Hybrid &53.40  &71.44  &42.08 &52.79 &35.12  &\textbf{40.70}  &33.25  &\textbf{36.61}  \\
{\:} + SKP &43.93  &63.02  &37.11 &46.61 &28.54  & 33.48 & 27.04 &29.86\\\midrule
{\:} + Two Tower& 38.51 & 58.23 & 33.54&42.44 &27.38  &32.57  &26.12  &29.01  \\
 \rowcolor{mygray} 
{\:} + {\ours} & \textbf{61.90} & \textbf{77.57}  &\textbf{47.44}  &\textbf{58.76}  & \textbf{44.27} & 15.41 & \textbf{44.27} & \textbf{22.93} \\
\midrule
\midrule
Flan-T5-xl & 10.86 & 19.16 & 10.40 &13.41 & 12.22 & 16.94 & 12.22 &14.21 \\
{\:} + G-Retriever &21.41  &39.37  &20.49 &26.91& 17.97 & 22.51 & 17.62 &19.76 \\
{\:} + G-RAG &19.72  &35.20  &18.86 &24.32 & 16.89 & 20.31 & 16.40 &18.17 \\
{\:} + Hybrid & 31.37 & 49.82 &27.67 &35.54 & 25.14 & 29.94 & 24.24 &26.77 \\
{\:} + SKP & 21.29 & 37.04 &20.53 &26.45 &18.82  &22.66  &18.32  &20.26  \\\midrule
{\:} + Two Tower&18.00  &33.05  &17.58 &22.83 &16.23  &20.22  &15.90  &17.81  \\
 \rowcolor{mygray} 
{\:} + {\ours} &\textbf{39.48}  &\textbf{56.94}  &\textbf{35.23}  &\textbf{43.51}  & \textbf{28.20} & \textbf{38.54} & \textbf{28.20} & \textbf{32.47} \\
\bottomrule
\end{tabular}
}
\vspace{-12pt}
\end{table*}

\textbf{Problem Formulation.}
We consider the graph retrieval-augmented generation (RAG) setting, where a large language model (LLM) generates answers based on a question \( q \) and a retrieved knowledge subgraph \( \mathcal{S}_q \). The model takes the pair \( (q, \mathcal{S}_q) \) as input, where \( \mathcal{S}_q \) is retrieved from a knowledge graph \( \mathcal{G} \) conditioned on \( q \), i.e., \( \mathcal{S}_q = \mathbb{Q}(q, \mathcal{G}) \), and \( \mathbb{Q} \) denotes the retriever. We define \( \mathcal{S}_q \) as the union of triplets \( \tau = (h, r, t) \), where each \( \tau \in \mathcal{G} \) represents a factual statement relevant to the question.

Following prior work~\cite{li2024simple}, we reduce subgraph retrieval to a triplet ranking task, where the goal is to learn a retriever \( \mathbb{Q}\) that assigns higher scores to relevant triplets \( \tau \) given the question \( q \). The retriever is optimized to improve downstream generation quality by supplying more informative context to the LLM.



\textbf{Establishing Question-triplet Alignment via Graph Augmentation.} In typical retrieval training, the retriever \( \mathbb{Q} \) is optimized to align questions with their corresponding documents. In our formulation, this translates to learning a mapping between a natural language question \( q \) and a relevant set of knowledge triplets \( \mathcal{T}_q \subseteq \mathcal{G} \). However, under the general RAG framework, only question-answer pairs are available, with no explicit supervision for question-triplet alignment~\cite{peng2024graph}. This motivates us to establish the question-triplet alignment by augmenting the knowledge graph. Specifically, we generate synthetic natural language questions from triplets by performing masked triplet prompting. For each triplet \( \tau = (h, r, t) \in \mathcal{G} \), we mask one entity to construct masked triplet \( \tau' \in \{([MASK], r, t), (h, r, [MASK])\} \), and prompt LLaMA-3.1-8B-Instruct~\citep{grattafiori2024llama} with instructions to generate a synthetic question in natural language $q_\tau$, corresponding to the original triplet $\tau$. 



Each synthetic question \( q_\tau \) aligns with its original triplet \( \tau \), as well as neighboring triplets \( \tau_{\text{nb}} \) that share at least one entity with \( \tau \). We treat both types as positive signals: the original triplet \( \tau \) is directly relevant to the question \( q_\tau \), while its neighbors provide contextual support for better understanding. Formally, we construct the positive set:
\begin{equation}
    \mathcal{D}_{\text{pos}} = \{(q_\tau, \tau),\; (q_\tau, \tau_{\text{nb}})\}, \tau_{\text{nb}} \in \mathcal{G},\; \tau_{\text{nb}} \cap \tau \neq \emptyset.
\end{equation}

We further introduce negative set $\mathcal{D}_{neg}$ by randomly sampling triplets from \( \mathcal{G} \) such that they do not overlap with \( \tau \), serving as irrelevant distractors $\tau_{\text{neg}}$:
\begin{equation}
    \mathcal{D}_{\text{neg}} = \{(q_\tau, \tau_{\text{neg}})\}, \tau_{\text{neg}} \cap \tau = \emptyset.
\end{equation}

By integrating these sets, we construct the final pretraining dataset that captures varying levels of alignment between queries and knowledge triplets:
\begin{equation}
    \mathcal{D} = \{(q_\tau, \tau, \tau_{\text{nb}}, \tau_{\text{neg}})\}.
\end{equation}

Examples of the graph augmentation procedure are available in Appendix~\ref{app:example}. 

\textbf{Mining Question-triplet Alignment with Pretraining.} To model the varying level of alignment constructed in dataset $\mathcal{D}$ with our retriever $\mathbb{Q}$, we firstly encode query and triplets with retriever $\mathbb{Q}$, then optimize the retriever with a structure-aware objective function.

\textit{\uline{Encoding.}} We adopt a two-tower architecture~\cite{karpukhin2020dense}, commonly used in information retrieval, as the basis for our retriever. It consists of separate encoders for natural language queries and knowledge triplets, denoted by \( E_q \) and \( E_\tau \), respectively, i.e., \( \mathbb{Q} = \{E_q, E_\tau\}\). For question $q_{\tau}$ and triplets $\tau$, $\tau_{nb}$, $\tau_{neg}$, their embeddings are computed as $z_q = E_q(q_{\tau})$, $z_\tau = E_\tau(\tau)$, $z_{nb} = E_\tau(\tau_{nb})$, and $z_{neg} = E_\tau(\tau_{neg})$, serving as a prerequisite for subsequent structure-aware optimization.   


\textit{\uline{Optimization.}} To effectively answer knowledge-intensive questions, the retriever should prioritize facts that are directly relevant to the query. In addition, supporting knowledge that addresses secondary aspects of the queried fact can provide helpful context and improve answer quality. On the other hand, retrieving irrelevant knowledge offers little benefit and may introduce noise, reducing overall performance. Based on this motivation, we optimize the retriever using a customized variant of triplet loss~\cite{schroff2015facenet}, which is designed to learn from varying levels of preference. Given a query \( q \), a more preferred triplet \( p \), and a less preferred triplet \( n \), the basic triplet loss is defined as:
\begin{equation}
    \mathcal{M}(p, n, q, \gamma) = \max(0, \gamma + \cos(n, q) - \cos(p, q)),
\end{equation}
where \( \cos(\cdot, \cdot) \) denotes cosine similarity, and \( \gamma \) is the margin hyperparameter. This formulation encourages the model to assign higher similarity to preferred triplets relative to less preferred ones.


We apply and customize this loss by enforcing a soft preference ordering among triplets conditioned on the question \( q_{\tau} \): the exact matching triplet \( \tau \) is preferred over its neighbors \( q_{\text{nb}} \), which in turn are preferred over irrelevant triplets \( q_{\text{neg}} \). The final pretraining loss is represented as:
\begin{equation}
    \label{eq:final}
    \mathcal{L} = \mathcal{M}(z_{\tau}, z_{nb}, z_q, \gamma_{1}) + \mathcal{M}(z_{nb}, z_{neg}, z_q, \gamma_{2}),
\end{equation}
with separate margins $\gamma_{1}$ and $\gamma_{2}$ for each preference level.



\begin{figure*}[h]
\centering

\caption{Top-ranked triplets retrieved by pretraining-free two tower retriever and {\ours}, given the query \textit{"What does Jamaican people speak?".}}
\vspace{-5pt}
\begin{minipage}[h]{0.49\textwidth}
\centering
\begin{tcolorbox}[title=Two Tower, colback=white, colframe=blue!50!black, fonttitle=\bfseries, left=.5pt, right=.5pt, top=1pt, bottom=1pt]
\scriptsize
\texttt{Kemar Bailey-Cole | languages | English Language} \\
\texttt{The problem of freedom | subjects | Jamaica} \\
\texttt{The Blue Lagoon | language | English Language} \\
\texttt{Hansle Parchment | nationality | Jamaica}
\end{tcolorbox}
\end{minipage}
\begin{minipage}[h]{0.49\textwidth}
\centering
\begin{tcolorbox}[title={\ours}, colback=white, colframe=blue!50!black, fonttitle=\bfseries, fontupper=\tiny, left=.5pt, right=.5pt, top=1pt, bottom=1pt]
\scriptsize
\texttt{Jamaican English | countries\_spoken\_in | Jamaica} \\
\texttt{Jamaican Creole English  | countries\_spoken\_in | Jamaica} \\
\texttt{Jamaica | languages\_spoken | Jamaican English} \\
\texttt{Jamaica | languages\_spoken | Jamaican Creole English} 
\end{tcolorbox}
\end{minipage}
\label{fig:case_study}
\vspace{-12pt}
\end{figure*}

\textbf{Leveraging Question-triplet Alignment during Inference.}
At inference time, the pretrained retriever \( \mathbb{Q} \), optimized with augmented question-triplet alignment with structure-awareness (Eq.~\ref{eq:final}), is used to query the knowledge graph \( \mathcal{G} \) and retrieve top-K triplets to construct the subgraph \( \mathcal{S}_q \). This subgraph provides knowledge-rich context to enhance the LLM’s performance, without requiring any additional fine-tuning on the question answering task.



\section{Experiments}
\label{sec:experiment}

\textbf{Experiment Settings.}
All experiments settings are available in Appendix~\ref{app:setting}.

\noindent\textbf{Evaluation Results.}
The experiment results are available in Table~\ref{tab:main_experiments}. Key insights we obtain through the analysis include:

\textit{\uline{Pretrained language models struggle to perform effective retrieval over structured graph data.}} Using language models pretrained on plain text as retrievers often results in limited and inconsistent gains. While effective for natural language understanding, they lack the structural alignment and graph-specific inductive biases necessary for reasoning over knowledge graphs. Notably, the two tower retriever, which shares the same architecture as {\ours} but omits graph pretraining, performs poorly, highlighting the limitations of relying solely on plain-text pretrained models for graph-based retrieval tasks.

\textit{\uline{{\ours} advances graph retrieval by bridging text and graph with structure-awareness.}} Across all evaluated settings, {\ours} consistently outperforms retrievers without targeted pretraining, effectively aligning textual queries with relevant subgraphs. Its improvements on downstream question answering tasks reflect the quality of retrieved subgraphs in providing accurate and contextually relevant information. Given that the pretraining-free two tower variant performs poorly, the strong performance of {\ours} stems not from its structure alone but from its carefully designed pretraining strategy. This highlights the effectiveness of our question-triplet alignment objective and our success in modeling structural relations between text and graph through pretraining.

\begin{figure}[t]
    \centering
        \caption{Performance vs K in the selection of top-ranked triplets.}
        \vspace{-5pt}
    \includegraphics[width=0.95\linewidth]{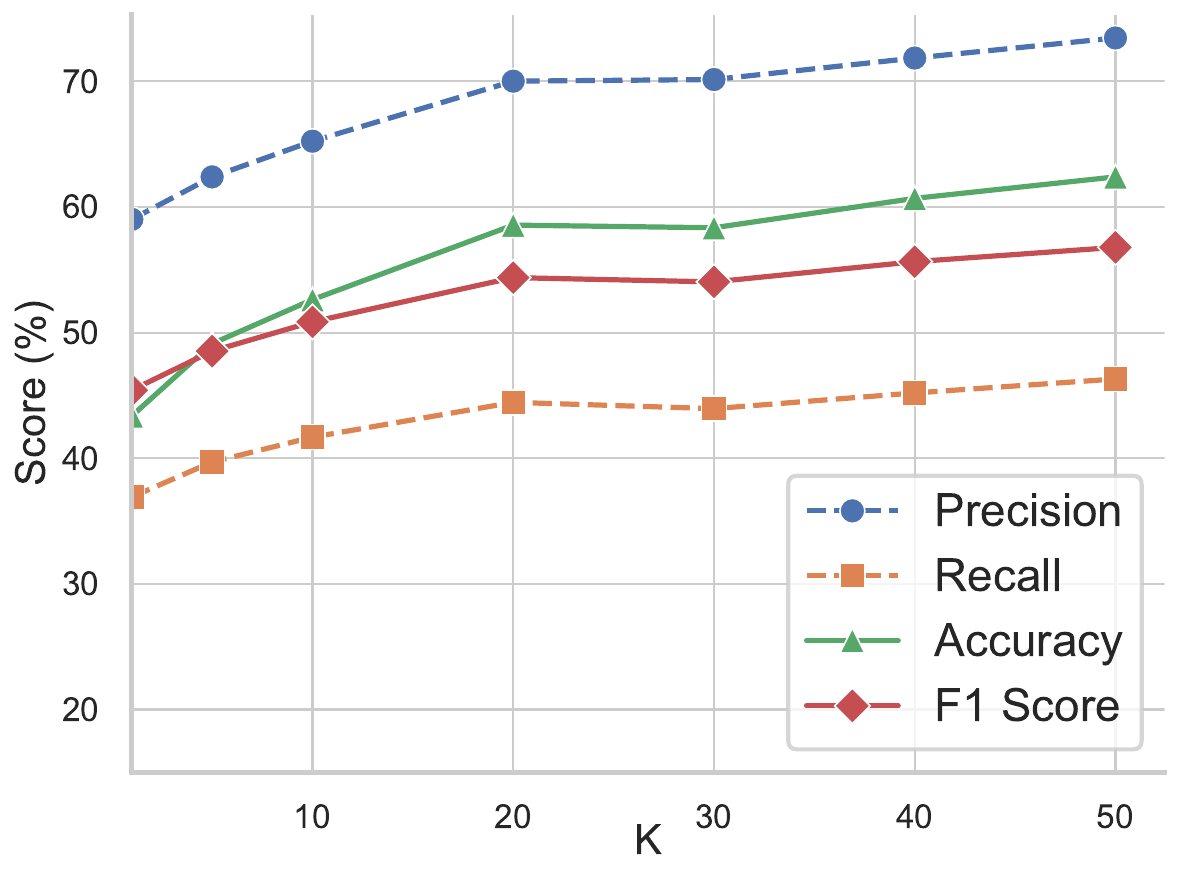}
    \label{fig:K_study}
    \vspace{-20pt}
\end{figure}


\textbf{Qualitative Analysis.} We conduct a case study to qualitatively analyze the retrieved facts of {\ours}, with results shown in Figure~\ref{fig:case_study}. Intuitively, methods relying on language models pretrained on general text struggle to bridge the gap between natural language queries and knowledge graph content, retrieving low-relevance and noisy results. In contrast, {\ours} benefits from knowledge graph-based pretraining with a discriminative optimization objective, resulting in retrieved facts that are more relevant and coherent with the input query.

\textbf{Quantitative Analysis.} We further conduct a quantitative analysis by varying the top-$K$ value used during triplet retrieval, as shown in Figure~\ref{fig:K_study}. All metrics exhibit a consistent upward trend as $K$ increases. These results highlight that {\ours} consistently benefits from an increasing amount of retrieved content, demonstrating its ability to capture broader context while maintaining robustness to potential noise introduced by retrieval.

\section{Conclusion}
\label{sec:conclusion}

We introduced {\ours}, a simple yet effective retriever pretrained on knowledge graphs to support retrieval-augmented generation over knowledge graphs. Through LLM-guided graph augmentation and structure-aware pretraining, {\ours} learns to align questions with informative subgraphs in a flexible and data-agnostic manner. Comprehensive experiments show that {\ours} consistently enhances both retrieval quality and downstream generation performance.

\clearpage
\section*{Limitations}

While {\ours} demonstrates strong performance in graph retrieval, it still has two limitations. First, our pretraining currently considers only 1-hop neighbors due to computational constraints, which may limit the model’s effectiveness in capturing larger contextual subgraphs or longer reasoning paths. Extending the method to incorporate multi-hop structures remains feasible and is worth exploring. Second, although the pretraining strategy is broadly applicable, we adopt a basic two-tower retriever for implementation due to limited bandwidth. Investigating more expressive retriever architectures presents a promising direction for future work.
\bibliography{custom.bib}

\appendix



\newpage
\section{Example of Graph Augmentation}
\label{app:example}
Table~\ref{tab:example} presents examples in Freebase~\cite{bollacker2008freebase} that appear in our augmented dataset, demonstrating how the graph augmentation process operates.
\begin{table*}[t]  
\centering
\renewcommand{\arraystretch}{1.2}
\begin{tabularx}{\textwidth}{|l|X|}
\hline
\textbf{Original Triplet} $\tau$ & (Attention deficit hyperactivity disorder, treatments, Modafinil) \\
\hline
\multirow{2}{*}{\textbf{Generated Questions $q_{\tau}$}} 
& \textit{What is a treatment for Attention deficit hyperactivity disorder?} \\
& \textit{What is Modafinil used to treat?} \\
\hline
\multirow{2}{*}{\textbf{Neighbor Triplets} $\tau_{nb}$} 
& (Attention deficit hyperactivity disorder, treatments, Zooey Deschanel) \\
& (Cephalon, product, Modafinil) \\
\hline
\multirow{2}{*}{\textbf{Negative Triplets} $\tau_{neg}$} 
& (Prednisone, active\_moiety\_of\_formulation, Prednisone 10 tablet) \\
& (Welcome To The Jungle, written\_by, Jonathan Hensleigh) \\
\hline
\end{tabularx}
\caption{Example of graph augmentation.}
\label{tab:example}
\end{table*}

\section{Experiment Details}
\label{app:setting}

\textbf{Datasets.}
Following previous studies\cite{luo2023reasoning}, we adopt two prevalent datasets for experiments, i.e., WebQSP~\cite{yih2016value}, under the CC BY 4.0 License, and CWQ~\cite{talmor2018web}, under the Apache-2.0 License. The WebQSP test set used for inference contains 1.628 question–answer pairs, while the CWQ test set comprises 3.531 pairs.

\textbf{Implementation.}
We implement our two-tower retriever with two distilbert-base-uncased~\cite{Sanh2019DistilBERTAD} encoders. We choose the same text encoder for all retrievers for the fair comparison. For methods requiring pretraining (SKP and {\ours}), we perform pretraining on a subset of Freebase~\cite{bollacker2008freebase} that includes entities related to the WebQSP and CWQ datasets, which are independent of the question answering task, eliminating any data leakage concern. Pretraining is conducted for 5 epochs using AdamW~\cite{loshchilov2017decoupled}, with a batch size of 512 and a learning rate of 2e-5. Margins $\gamma_1$ and $\gamma_2$ in Eq.~\ref{eq:final} are both set to 0.5. The retrievers are further evaluated in the zero-shot setting.

\textbf{Backbones.}
Retrieval strategies are evaluated with LLM backbones pretrained in general domains, including ChatGPT-3.5 Turbo~\cite{achiam2023gpt}, LLaMA2-7B~\cite{touvron2023llama}, and Flan-T5-XL~\cite{chung2024scaling}, covering large language models of varying sizes and both open- and closed-source types. The usage of these artifacts aligns with their intended use for research purposes.  

\textbf{Computational Devices.} All experiments were conducted on four NVIDIA A6000 GPUs with CUDA version 12.0, running on an Ubuntu 20.04.6 LTS server. 

\textbf{Baselines.} 
We include the following graph-retrieval baselines for comparison:

\uline{\textit{G-Retriever}}~\cite{he2024g} is a retrieval-augmented generation framework designed for question answering over textual graphs. It retrieves relevant nodes and edges based on semantic similarity and constructs subgraphs using the Prize-Collecting Steiner Tree (PCST) algorithm to form concise, query-relevant subgraphs for generation .

\uline{\textit{G-RAG}}~\cite{hu2024grag} is a graph retrieval-augmented generation method that enhances LLMs by retrieving and integrating textual subgraphs. It represents subgraphs as pooled embeddings of k-hop ego-graphs and retrieves them to incorporate both textual and topological information through dual prompting, improving performance on multi-hop reasoning tasks .

\uline{\textit{Hybrid}}~\cite{li2023graph} is a hybrid retrieval model that combines sparse retrieval (BM25) and dense retrieval (DPR) for coarse retrieval, followed by reranking with a cross-encoder to improve retrieval performance.

\uline{\textit{SKP}}~\cite{dong2023bridging} leverages traditional approaches like contrastive learning and masked language prediction on graphs to obtain a more graph-concentrated encoder for retrieval, enhancing the model's ability to represent complex subgraphs.

\uline{\textit{Two-tower}}~\cite{karpukhin2020dense} is a dense passage retrieval approach for open-domain question answering. Utilizing a dual-encoder framework, it learns dense representations from question-passage pairs, outperforming traditional sparse retrieval methods like BM25 in top-20 passage retrieval accuracy.

All the baselines are required to retrieve knowledge without prior information about entities in question or answer of the question. For approaches containing multiple stages such as GNN-tuning~\cite{he2024g, hu2024grag} or parameter-efficient fine-tuning~\cite{hu2024grag}, we just take their PLM-based graph-retrieval module for fair comparison.

\section{Potential Risk}

Although {\ours} demonstrates strong performance, it is still possible for the retrieved results to reflect biases. Blind reliance on these results—treating them as factual without contextual verification—may raise societal concerns. Users of {\ours} are encouraged to critically assess the retrieved content within the specific application context to mitigate potential ethical risks.

\end{document}